\documentclass[aip,apl, reprint,superscriptaddress, nobibnotes, nofootinbib]{revtex4-1}
\usepackage{graphicx}
\usepackage{color}
\usepackage[dvipsnames]{xcolor}
\usepackage[normalem]{ulem}
\usepackage{float}
\usepackage{amssymb}
\usepackage{soul}
\usepackage{hyperref}

\newcommand{\blfootnote}[1]{
  \begingroup
  \renewcommand\thefootnote{}\footnote{#1}
  \addtocounter{footnote}{-1}%
  \endgroup
}
\newcommand{\LLaffil}{\affiliation{Lincoln Laboratory, Massachusetts Institute of Technology, 244 Wood Street, Lexington, MA 02421}}
\newcommand{\RLEaffil}{\affiliation{Research Laboratory of Electronics, Massachusetts Institute of Technology, 77 Massachusetts Avenue, Cambridge, MA 02139}}
\newcommand{\Physaffil}{\affiliation{Department of Physics, Massachusetts Institute of Technology, Cambridge, MA 02139}}
\newcommand{\EECSaffil}{\affiliation{Department of Electrical Engineering and Computer Science,
Massachusetts Institute of Technology, Cambridge, MA 02139}}

\begin{document}

\title{Characterization of superconducting through-silicon vias as capacitive elements in quantum circuits}

\author{T. M. Hazard} \LLaffil
\author{W. Woods} \LLaffil
\author{D. Rosenberg} \LLaffil
\author{R. Das} \LLaffil
\author{C. F. Hirjibehedin} \LLaffil
\author{D. K. Kim} \LLaffil
\author{J. M. Knecht} \LLaffil
\author{J. Mallek} \LLaffil
\author{A. Melville} \LLaffil
\author{B. M. Niedzielski} \LLaffil
\author{K. Serniak} \LLaffil \RLEaffil
\author{K. M. Sliwa} \LLaffil
\author{D. R. W. Yost} \LLaffil
\author{J. L. Yoder} \LLaffil
\author{W. D. Oliver} \LLaffil \RLEaffil \Physaffil \EECSaffil
\author{M. E. Schwartz} \LLaffil
\date{\today}

\begin{abstract}
The large physical size of superconducting qubits and their associated on-chip control structures presents a practical challenge towards building a large-scale quantum computer.  In particular, transmons require a high-quality-factor shunting capacitance that is typically achieved by using a large coplanar capacitor.  Other components, such as superconducting microwave resonators used for qubit state readout, are typically constructed from coplanar waveguides which are millimeters in length.  Here we use compact superconducting through-silicon vias to realize lumped element capacitors in both qubits and readout resonators to significantly reduce the on-chip footprint of both of these circuit elements.  We measure two types of devices to show that TSVs are of sufficient quality to be used as capacitive circuit elements and provide a significant reductions in size over existing approaches.

\end{abstract}

\blfootnote{For correspondence: thomas.hazard@ll.mit.edu,\\mollie.schwartz@ll.mit.edu}

\maketitle

%\section{Introduction}

Superconducting qubits, a promising hardware platform for quantum computing applications, have progressed from proof-of-principle single-qubit demonstrations to mature deployments of many-qubit quantum processors \cite{Jurcevic2021,Chen2021,Krinner2022}. With increased processor size comes the need for vertical integration, which previously motivated the development of high-density superconducting through-silicon vias (TSVs) for connecting grounds and routing signals between the top and bottom sides of a chip \cite{Vahidpour2017,Yost2020,Mallek2021}. However, the potential utility of a TSV goes far beyond signal routing. In particular, a high-aspect-ratio TSV can have large self-capacitance that provides a means to miniaturize the largest lateral footprint components of superconducting quantum processors: the coplanar capacitance of the qubits and resonators. In this work, we demonstrate a compact TSV-enabled lumped-element resonator that provides vertical readout integration in a footprint smaller than that of a standard transmon qubit. We also demonstrate high-coherence transmon qubits for which TSVs provide the majority of the capacitance across the Josephson junction, shrinking the on-chip footprint of the qubit by a factor of approximately 30. TSV integration thus provides a remarkable new method for scaling and improving superconducting quantum processors.

Previously, TSVs have been utilized to route signals to planar qubits \cite{Yost2020} or to shift electromagnetic modes in substrates to higher frequencies\cite{Vahidpour2017}. While these are important applications of the technology, TSVs can have a much broader impact on superconducting qubits. In particular, they have the potential to be integrated into qubit circuits in a fundamentally different manner, in which their geometry is exploited to construct novel 3D devices that are not possible in a 2D architecture. The high capacitance density enabled by TSVs can reduce the size of certain large qubit control and readout elements, allowing for a more compact tiling of qubits and their associated electronics. Furthermore, TSVs can be used to provide the large shunting capacitance needed for qubits such as transmons\cite{Koch2007}, C-shunt flux qubits\cite{yan2016flux} or $0-\pi$ qubits\cite{gyenis2021}, eliminating the requirement for a large area coplanar capacitor.  We present the design and characterization of a novel TSV-integrated compact resonator for qubit readout and a compact transmon qubit incorporating TSVs. Our results, which represent the first use of superconducting TSVs as capacitive elements in a qubit circuit, establish the feasibility of using superconducting 3D technologies to enable new qubit and control circuit geometries. 

\begin{figure*}[t]
    \centering
    \includegraphics{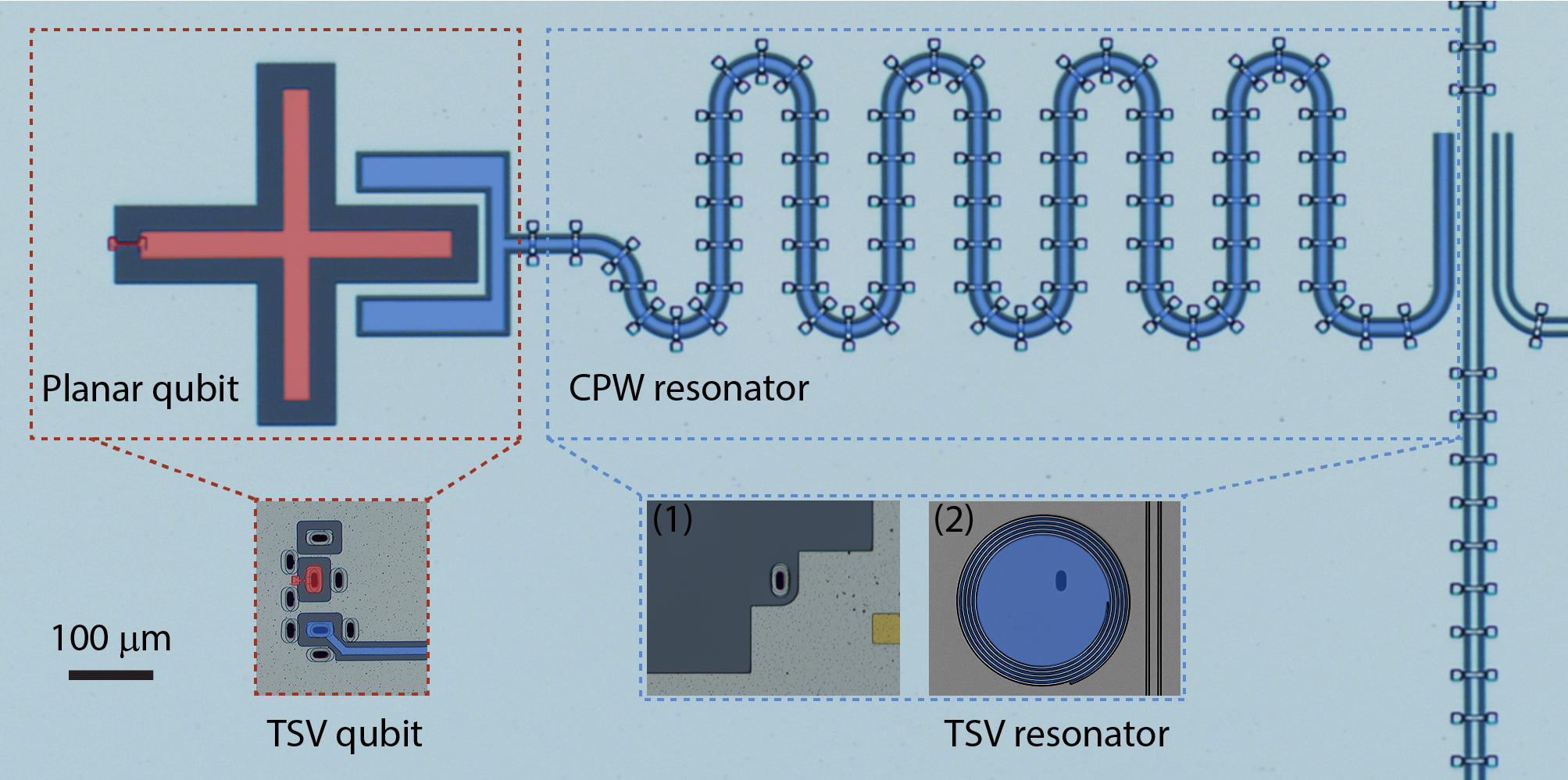}
    \caption{An optical micrograph of a planar superconducting qubit-resonator system.  The device consists of a transmon qubit (highlighted in red) with a single Josephson junction and the large capacitive shunt which are coupled to a CPW resonator (highlighted in blue).  The insets are images, taken at the same scale, of the qubit and resonator components from two devices which are miniaturized by using TSVs (the dark ovals in the images) for the capacitive circuit elements.  The lumped element resonators are on a separate chip and are bump bonded to a qubit chip, with the side (1) facing the qubit and coupled via the TSV resonator going through the chip to side (2).}
    \label{fig:Schematic}
\end{figure*}

%\section{Results}
%\subsection{Compact TSV Readout Resonators}
One of the obstacles to building addressable qubit arrays is the size of each qubit's associated control and readout elements, which can be significantly greater than the inter-qubit spacing. In a typical configuration, a superconducting resonator is constructed from meandering sections of a coplanar waveguide (CPW) transmission line, terminated on each end of the resonator with either an open or a short circuit to ground.  The termination determines the fundamental mode of the resonator, either $\lambda/4$ or $\lambda/2$ where $\lambda$ is the mode wavelength\cite{Pozar:882338}.  For readout multiplexing, several resonators are coupled to a shared transmission line\cite{Zmuidzinas2012}, with one end near the transmission line and the other end near the qubit.  In the limit of weak external coupling and high internal quality factors, the resonator frequency is set by the inductance and capacitance per unit length, $l$ and $c$, respectively, along with the total resonator length\cite{Goppl2008}.  While using a CPW resonator is straightforward from a design and fabrication perspective, these resonators occupy a large area, i.e., the CPW has a length of order $\lambda$ which corresponds to several mm for mode frequencies in the 6-8 GHz range (see Fig. \ref{fig:Schematic}). Additionally, higher harmonics of the fundamental modes can also interact with the qubit, causing enhanced relaxation when the qubits are higher in frequency than the resonator\cite{Houck2008}.  Lumped-element resonators can offer an improvement in both size and higher mode issues (a lumped element resonator does not explicitly host higher harmonics at integer multiples of the fundamental mode frequency) in CPW resonators, but realizing a high capacitance density in planar elements has previously proved elusive.

\begin{figure}
    \centering
    \includegraphics{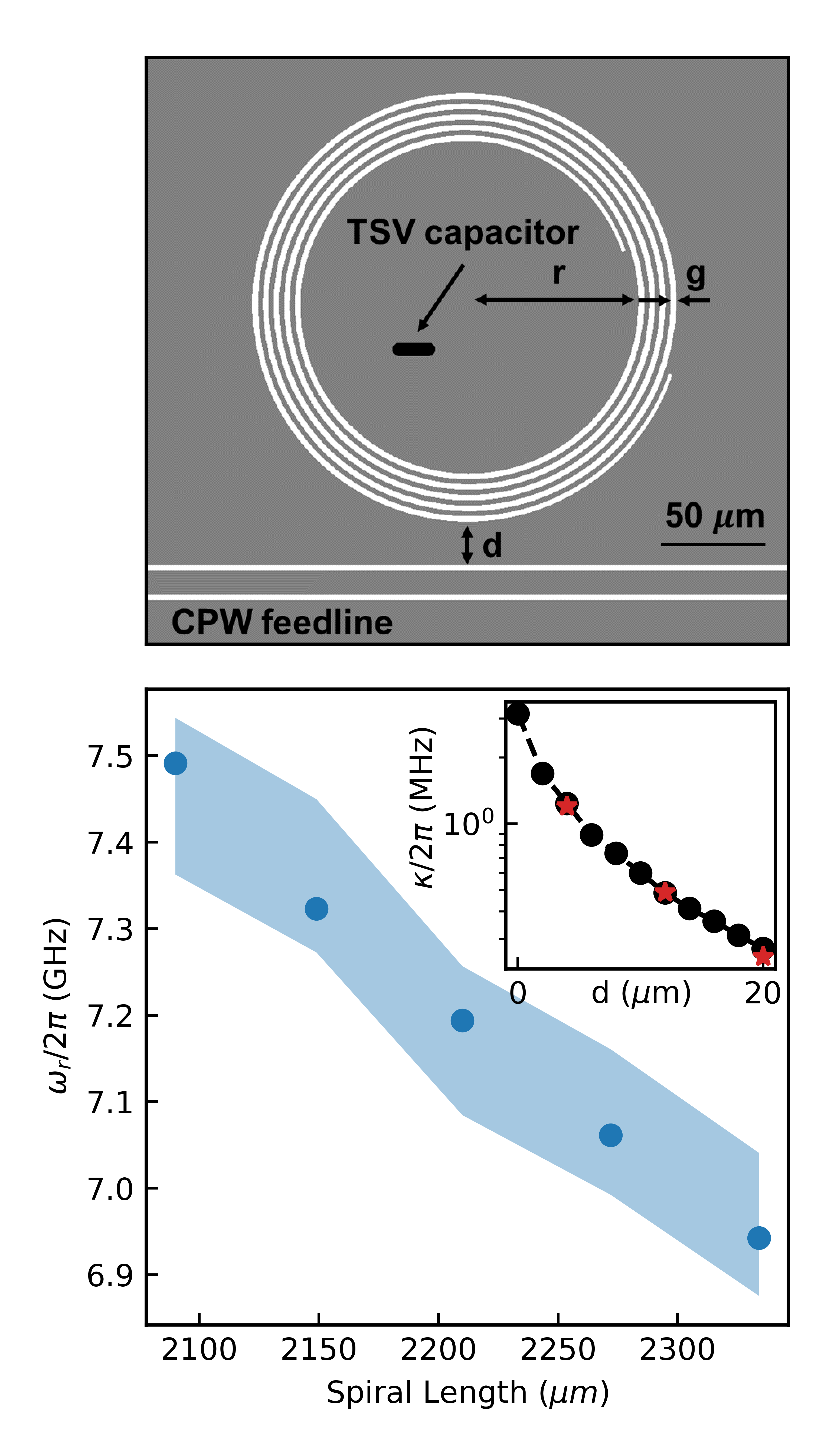}
    \caption{Top panel: A schematic view of the spiral resonator coupled to the input/output feedline.  Bottom panel: Measured resonator frequencies (circles) as a function of length of the spiral inductor.  The colored bands represent the range of simulated resonator frequencies using $L_k=$2.0-2.2 pH$/\square$, to account for $L_k$ variation across different chips across a wafer.  The inset shows the simulated (black circles) and measured (red stars) values of the resonator line widths, $\kappa$, for different values of the distance between the resonator and CPW feedline.}
    \label{fig:CircleResonators}
\end{figure}

To simultaneously realize a high quality, large capacitance and small form factor, we use a superconducting TSV integrated into a compact lumped-element resonator.  The resonator, shown in Fig \ref{fig:CircleResonators}, comprises a central disk (with radius $r$) connected to a TSV and a meandering spiral (with line width $w$) and gaps ($g$) between adjacent spirals. The offset distance ($d$) between the edge of the spiral and the feedline sets the coupling quality factor of the resonator, $Q_c$.  In all of the designs presented here, the resonators are over-coupled, with their internal quality factors $Q_i$ much greater than $Q_c$, which is related to the resonator linewidth $\kappa=Q_c/\omega_r$.  The resonators are capacitively coupled to the qubits via the TSV.  The central disk, meandering spiral, and TSV are all constructed from a 200-nm-thick superconducting TiN film.  Extended details of the TSV fabrication process are provided in previous work \cite{Mallek2021}.  From a set of test $\lambda/4$ CPW-resonator test structures, we extract a kinetic inductance $L_k\approx2.0$ pH/$\square$ of the TiN film.  The combination of a high kinetic inductance film and a laterally compact TSV capacitor allows for a lumped element resonator that is less than $180\ \mu$m in diameter, smaller in dimension than capacitors typically used in transmon qubits.  We note that, in principle, a lumped element resonator could be constructed with a Josephson junction array or a higher-$L_k$ material for even greater kinetic inductance density and, thereby, a smaller-sized resonator \cite{Grunhaupt2019,Pechenezhskiy2020}.  Although not fundamental to the operation of the device, the resonators and qubits are fabricated on two separate chips, with an interposer chip comprising the resonators, shielding/ground TSVs, flux bias lines, and microwave antennas, as well as a qubit chip with transmon qubits, superconducting bump bonds, and etched-silicon spacers to set the separation distance of 3 $\mu$m between two chips\cite{Niedzielski2019}.

The bump-bonded devices are measured in a dilution refrigerator with base temperature $T\approx12$ mK.  We find quantitative agreement between the measured resonator frequencies and those simulated using Sonnet electromagnetic solver ($f_\mathrm{sim}$), shown in Fig. \ref{fig:CircleResonators}.  We vary the number of turns of the spiral inductor and compare the measured and simulated frequencies (with $L_k=2$ pH$/\square$) and find an average difference of $\bar{f}_\mathrm{diff}=$ 80 MHz or $\bar{f}_\mathrm{diff}/f_\mathrm{sim}=$ 1\%.  In addition to the spiral length variations, several additional devices were tested with reduced offset distance $d$ to increase $\kappa$ for faster qubit readout (inset of Fig. \ref{fig:CircleResonators}).

\begin{figure}
    \centering
    \includegraphics{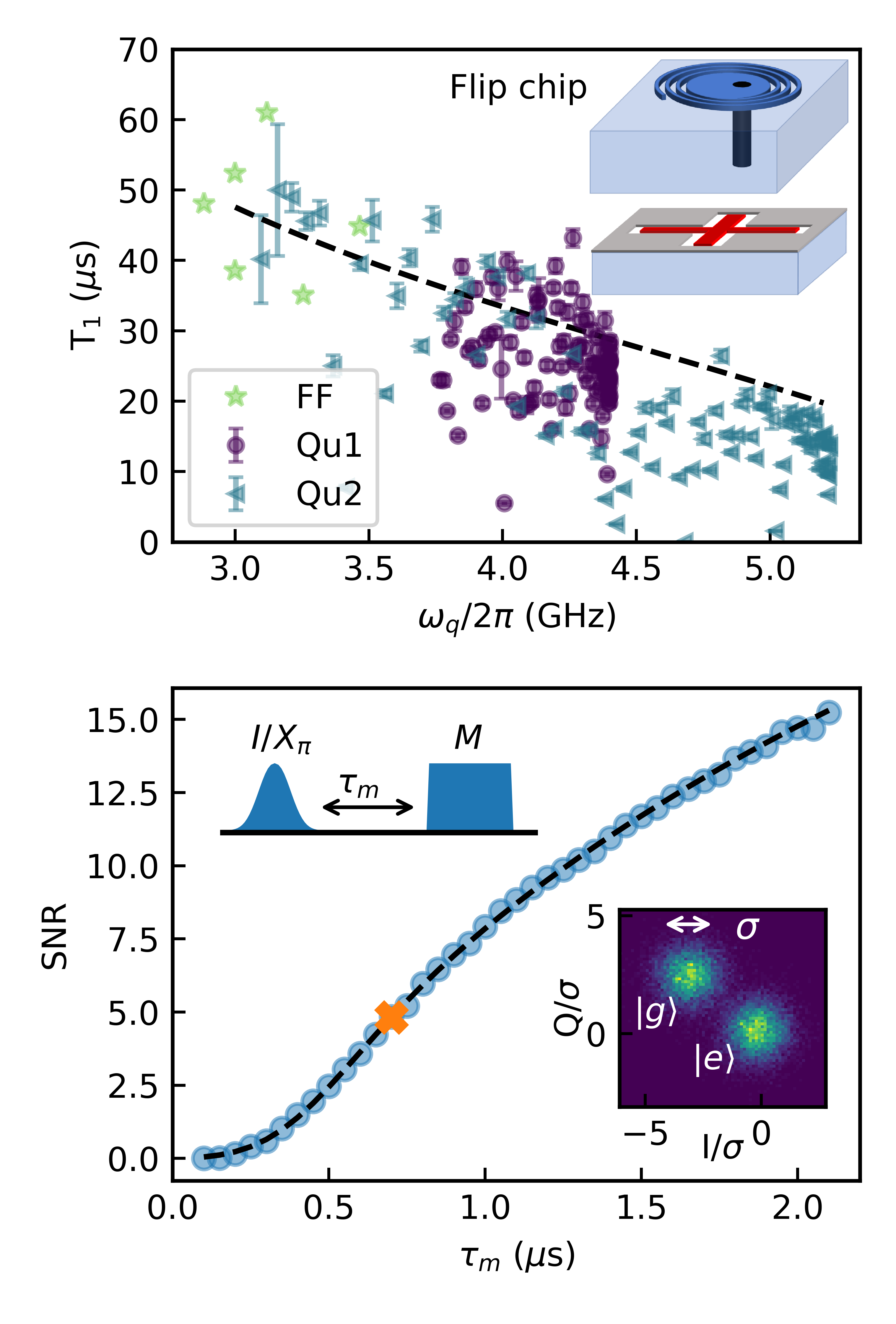}
    \caption{Top panel: Measurement of transmons bump bonded to the readout resonators on the TSV chip (schematic inset).  We plot the $T_1$ for two tunable transmons (circles and triangles) as well as several single junction fixed frequency qubits (stars). The dashed  line is a fit of $T_1$ vs $\omega_q$ accounting for dielectric loss and Purcell relaxation.  Bottom panel: Readout characterization of the flux tunable qubit device, ``Qu1", showing SNR vs $\tau_m$. The dashed black line is a fit to Eqn. \ref{eqn:SNR}.  Inset: Histograms of the I and Q voltage quadratures for $\tau_m=700$ ns (orange cross for SNR=5), with the qubit prepared in either the ground or excited state.  The readout voltages are normalized by the width, $\sigma$, of the voltage distributions of the $10^4$ trials.}
    \label{fig:TwoTier}
\end{figure}

We measure the coherence times of several transmon qubits to further characterize the coupled qubit and resonator system. The qubit chips consist of both fixed frequency, single Josephson junction (JJ), transmon qubits and flux-tunable transmons with two JJs placed in a SQUID loop.  Changing magnetic flux through the SQUID loop changes the qubit frequency, proportional to $\propto \sqrt{2E_{J}|\cos\left(\pi\Phi/\Phi_0\right)|}$, where $E_{J}$ is the Josephson energy of the two JJs (designed here to be equal), $\Phi$ is the flux through the SQUID loop, and $\Phi_0$ is the magnetic flux quantum.  The values of measured energy relaxation time $T_1$ for both the tunable and fixed-frequency transmons are shown in Fig. \ref{fig:TwoTier}.  Additionally, we find an echoed coherence time $T_{2E}$ with average value $T_{2E}=47\ \mu$s for the fixed frequency transmons and $T_{2E}\approx2T_1$ at $\Phi=0$ (flux-sweet spot) for the flux tunable transmons.  The black dashed line is the expected $T_1$ vs $\omega_q$ accounting for both dielectric loss and Purcell relaxation\cite{Houck2008} through the readout resonator. For a frequency-independent dielectric quality factor $Q_\mathrm{diel}=\omega_q T_1$\cite{Wang2015}, we find $Q_\mathrm{diel}=10^6$, which is consistent with the value we measure in other flip-chip devices coupled to CPW resonators.  For the tunable transmon labeled ``Qu1," we measure a dispersive shift $2\chi=930$ kHz of the resonator at the qubit maximum frequency near 4.45 GHz.  From decay measurements of the output voltage of the resonator when the readout drive is turned off, we extract $1/\kappa=$ 300 ns.

To characterize the readout visibility of the TSV resonator, the qubit is prepared in either its ground or excited state and then measured.  The preparation and measurement sequence is repeated $10^4$ times for each state and the resulting complex readout voltages are histogrammed (Fig. \ref{fig:TwoTier}, bottom inset).  The distributions are fit to Gaussians of width, $\sigma$, and the voltages are then normalized by $\sigma$.  In this way, the signal-to-noise ratio (SNR) can be expressed as the separation of the normalized ground- and excited-state readout voltages.  This approach is not a measure of the full state preparation and measurement fidelity, which would include pulse errors, qubit relaxation, and thermal population of excited states, but rather a method to characterize the readout separation fidelity of the qubit ground and excited states \cite{Jeffrey2014}.  We characterize the SNR as a function of measurement time, shown in Fig. \ref{fig:TwoTier}.  In the dispersive regime and long measurement time limit, the SNR is given by
\begin{equation}
\mathrm{SNR} (\tau_m\rightarrow \infty)\approx\frac{2\epsilon}{\kappa}\sqrt{2\kappa\tau_m} |\sin2\phi|
\label{eqn:SNR}
\end{equation}
where $\tau_m$ is the measurement time, $\epsilon$ is the readout drive amplitude, and $\phi$ is the phase shift of the resonator between the two qubit states\cite{Didier2015,Blais2021}.  The separation fidelity can be expressed as $F_s=1-\mathrm{erfc}\left(\mathrm{SNR}/2\right)$, which for Qu1, exceeds 99.9\% for $t_m\ge 700$ ns.  In principle, even faster readout can be achieved with an increased resonator $\kappa$ (decreased distance $d$ to the feedline) as well as incorporating a TSV based compact filter to protect the qubit from relaxation due to the Purcell effect\cite{Reed2010,Jeffrey2014}.

%\subsection{Compact TSV Qubits}
At the current error rates in state-of-the-art qubits, a large-scale quantum computer based on superconducting qubits would likely need $10^5$ to $10^8$ qubits\cite{Fowler2012}.  While transmon qubits are one of the leading building blocks today, their physically large shunt capacitor makes scaling to thousands of qubits a daunting engineering challenge.  Additionally, at large chip sizes, low-quality-factor cavity modes associated with the chip enclosure can couple to qubits, causing a degradation in performance\cite{Sheldon2017}.  With these challenges in mind, several groups have sought to reduce qubit size via different means, including the use of interdigitated capacitors\cite{Weides2011}, or using the high specific capacitance of the junction itself\cite{Zhao2020,Mamin2021}.

\begin{figure}
    \centering
    \includegraphics{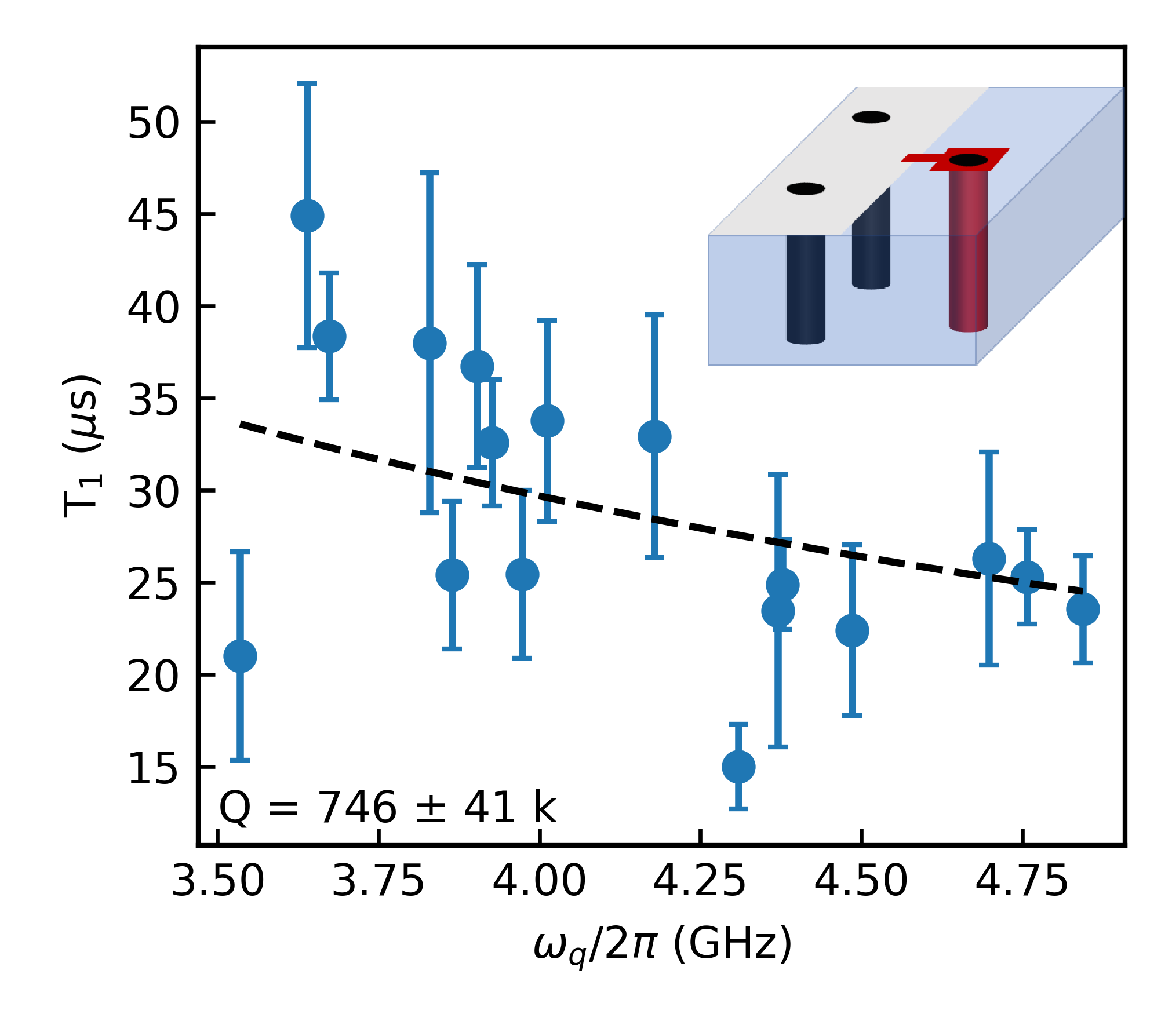}
    \caption{$T_1$ vs $\omega_q$ measured for the TSV transmon qubits (schematic inset with a single TSV for the qubit capacitor).  The black dashed line is a constant $Q_\mathrm{diel}$ fit. The error bars indicate the distribution of $T_1$'s measured for each of the qubits over the span of 24 hours.  The inset 3D schematic corresponds to the optical image of the TSV qubit in Fig. \ref{fig:Schematic}, in which the red and black cylinders correspond to the qubit and ground TSVs respectively.}
    \label{fig:SmallestQubit}
\end{figure}

To this end, we designed a set of transmon qubits with a Josephson junction connected to a TSV capacitor on a single chip, shown in the left inset of Fig. \ref{fig:Schematic}.  The device consists of the 10 x 20 $\mu$m TSV capacitor with a 20 $\mu$m gap to ground for a total lateral device size of 40 x 50 $\mu$m, an areal size reduction of a factor of 30 compared to the size of typical state-of-the-art transmon qubits \cite{Gordon2022}.  A series of these qubits where fabricated across a range of frequencies spanning 3.5-4.8 GHz, by changing the JJ area, i.e., the critical current and thereby the Josephson energy $E_J$.  The total shunt capacitance of the qubits was nominally fixed at $85$ fF, with the contribution from the TSV capacitor being approximately 93-98\% of the total capacitance.  The $T_1$ times for the qubits are shown in Fig. \ref{fig:SmallestQubit}, where the error bars give a measure of the temporal variation over the course of 24 hours.  The TSV itself provides the majority of the total shunt capacitance and therefore the qubit $T_1$ is a good indicator for the quality of the TSV capacitor.  We can again use a constant $Q_\mathrm{diel}$ vs $\omega_q$, to estimate a quality factor $Q_\mathrm{diel}=746\ \pm 41\times 10^3$  which is comparable to what has been measured in other small footprint capacitor based transmon qubits\cite{Zhao2020,Mamin2021}.  Previously measured values of dielectric loss in superconducting devices indicate that $Q_\mathrm{diel}= 4-5\times 10^6$ for bulk silicon substrates\cite{Woods2019}, suggesting that the lower $Q_\mathrm{diel}$ measured in the TSV qubits comes from the interface between the TiN lining the TSV and the silicon substrate.  A future study of different geometries, with varying amounts of TiN/silicon surface area, can help put a stricter quantitative bound on this loss.

%\section{Conclusions}
We have created superconducting qubits and lumped element readout resonators with integrated superconducting TSVs.  The reduced lateral size of the readout resonators enables tiling of qubit arrays at a pitch limited by the size of the coplanar-capacitor-based transmon qubits.  We additionally characterized the coherence of transmon qubits constructed with TSVs and find the quality comparable to what has been measured in other high-capacitance-density structures\cite{Zhao2020,Mamin2021}.  Future extensions of TSV-based superconducting devices include the integration of TSVs as Purcell filters\cite{Reed2010}, which would feature a similar size reduction as the readout resonators, as well as investigating the mechanisms which limit $Q_\mathrm{diel}$ in the TSV process to improve the $T_1$ in the TSV transmon qubits.  TSVs should also prove advantageous for qubits made of novel materials which may have increased microwave losses\cite{Hazard2022}.  While current flip chip technologies allow coupling for qubits between the surfaces of two chips\cite{Gold2021,kosen2022}, a fully 3D qubit which extends through a chip could be used to improve the feasibility of a 3D color code\cite{Bombin2006}, which has the potential advantage of needing only a single round of local measurement for error correction \cite{Bombin2015} compared to toric codes\cite{Kitaev1997}.  

%\section{Acknowledgement}
This material is based upon work supported by the U.S. Department of Energy, Office of Science, Office of Advanced Scientific Computing Research, High Performance Computing and Network Facilities (Rep - Quantum Testbeds), under contract number: FWP \#FP00008338, and by the Under Secretary of Defense for Research and Engineering under Air Force Contract No. FA8702-15-D-0001. Any opinions, findings, conclusions or recommendations expressed in this material are those of the author(s) and do not necessarily reflect the views of the Department of Energy or the Under Secretary of Defense for Research and Engineering.

\bibliography{references}
\end{document}